\documentclass[12pt,a4paper]{article}

\usepackage{amssymb,amsmath}

\usepackage{graphicx}

\newcommand{\ra}{\rightarrow}
\newcommand{\be}{\begin{equation}}
\newcommand{\ee}{\end{equation}}
\newcommand{\ba}{\begin{eqnarray}}
\newcommand{\ea}{\end{eqnarray}}

\begin{document}

{\bf \centerline{\Large Static spherically symmetric black holes with scalar field}}

\bigskip

\centerline{J. Tafel}

\bigskip
\noindent
\centerline{Institute of Theoretical Physics, Faculty of Physics,}
\centerline{ University of Warsaw,  Poland, email: tafel@fuw.edu.pl}

\begin{abstract}
Static spherically symmetric black holes and particle like  solutions with  self interacting minimally coupled scalar field $\varphi$ are analyzed. They are  asymptotically flat or anti-de Sitter (AdS). We express them in terms of a single function $\rho$ which undergoes  simple conditions. If $\varphi$ is nontrivial  the ADM  mass $M$  has to be positive. No-hair theorems are generalized to the AdS asymptotic. For both asymptotics the Killing horizon is nondegenerate and its radius  cannot be bigger than $2M$. Derivatives of $\rho$ at  singularity determine  properties of admissible potentials $V(\varphi)$ as regularity, boundedness and behaviour for maximal values of $\varphi$. Several classes of solutions with singular or nonsingular potentials are obtained. Their examples  are presented in a form of plots.
\end{abstract}


\section{Introduction}

The first solution of the Einstein equations with  scalar field $\varphi$ was found by Fisher \cite{f}. Then it was rediscovered by several authors \cite{s,bl,b,jnw}. Fisher's  solution is static, spherically symmetric and  $\varphi$ is massless. It generalizes  the Schwarzschild solution, but its global structure is completely different. If $\varphi$ is nontrivial   singularity and  the event horizon are at the same place \cite{jnw}.

Later studies on the Einstein-scalar equations   led  to a number of no-hair theorems on static spherically symmetric solutions representing black holes or particle like solutions which are asymptotically flat \cite{r,cha,bek,tei,ap,h,sud,bek2,gl,br1,bs}. They show that  in the  case of  black holes  potential $V$ of the scalar field cannot be  positive definite simultaneously at all points outside the event horizon. If there is no horizon and  $V$ is positive definite everywhere a naked singularity must be present. The global structure of the maximally extended metric  is that of the Schwarzschild or Minkowski spacetime. Some of these results were  generalized to asymptotically anti-de Sitter  or de Sitter solutions \cite{br1,bs} and more involved models of  gravity interacting with  scalar field, also in higher dimensions (see \cite{br} for a review). On the other hand examples of black holes and particle like solutions, which do not undergo the no-hair theorems,  were constructed  (see e.g. \cite{bs,tmn,ns,mtz,ao}). Since the standard energy conditions are often not respected in modern relativity, these solutions can  still be valuable.

In this paper we study exclusively  static spherically symmetric solutions of the Einstein equations with minimally coupled scalar field. They are assumed to be asymptotically flat or AdS. We consider two classes of fields: black holes with the regular  Killing horizon and no naked singularities  and  particle like solutions with no singularities at all. In the latter case it follows from the Einstein equations that there is no horizon.
In order to avoid numerical analysis we assume that  function  $V(\varphi)$ is not a'priori prescribed. In this case the Einstein equations form an underdetermined system of three equations for four unknowns: $\varphi$, $V$ and two metric coefficients (other two can be fixed by means of a coordinate transformation), all depending on a radial coordinate $r$. We solve these equations (section 2) and express   all fields  in terms of a free function $\rho$ which is monotonically growing, convex, has a zero point and  approximates  $r-3M$ when  $r$ is large.  Given $\rho(r)$  a dependence of the potential $V$ on $\varphi$ is defined in a parametric way. 

This description allows to obtain new results and refine old ones. We generalize the no-hair theorems to the AdS asymptotic. As in the asymptotically flat case potential $V$ cannot be positive definite outside the Killing horizon, or everywhere in the case of particle like solutions, if scalar field is nontrivial in this domain.(section 5 and 6). 
Other results refer to both asymptotics. Despite of violation of standard energy conditions (dominant, strong and weak) except the null one, for any potential $V(\varphi)$ the total ADM mass $M$ has to be positive or $\varphi$ is trivial and metric is flat or AdS (section 6). Singularity  is unavoidable in the case of black holes  \cite{br1} (see section 4 for a proof). It occurs at $r_0>0$ if  necessary properties of the function $\rho$ break down at $r_0$, otherwise it is at $r=0$.  There is only one horizon \cite{br1}. Its radius $r_h$ obeys the Penrose inequality $r_h\leq 2M$  and the surface gravity is never zero (section 3 and 5).

The representation of solutions  in terms of $\rho$ allows also to obtain general properties of admissible potentials (section 4 and 5). Potential $V$ and its derivative  $V_{,\varphi}$ have to vanish at $\varphi=0$.
In generic case $\varphi$ is finite at singularity and $V$ is infinite. Regular potentials $V(\varphi)$ are possible, bounded or unbounded, if singularity is at $r=0$ and  derivatives of $\rho$ at this point are properly chosen.  Examples of solutions are easy to construct thanks to the description of necessary and sufficient conditions on the function $\rho(r)$. They are plotted  in Figure 1.

  In section 2 asymptotically flat solutions of the Einstein-scalar equations are represented in terms of the free function $\rho$. In section 3 we study  maximal extensions of solutions admitting the nonsingular Killing horizon. Section 4 is mainly devoted to properties of potentials $V(\varphi)$ as functions of scalar field $\varphi$. Section 5 contains generalizations to solutions with the AdS asymptotic. Section 6 deals with particle like solutions with any of the two asymptotics. Main results are collected in Theorems 1-7 and Remarks 1-4.

\section{Solutions with undefined $V(\varphi)$}
 The Einstein equations with  scalar field $\varphi$, potential $V(\varphi)$ and the cosmological constant $\Lambda$ take the form
\be
 R_{\mu\nu}-\frac{1}{2}Rg_{\mu\nu}=\varphi_{,\mu}\varphi_{,\nu}+(\Lambda+V-\frac{1}{2}\varphi^{,\alpha}\varphi_{,\alpha})g_{\mu\nu}\ ,\label{1a}
 \ee
 where $R_{\mu\nu}$ is the Ricci tensor of metric $g_{\mu\nu}$ and units $c=1$, $8\pi G=1$ are chosen. If $\varphi_{,\mu}\neq 0$ it follows from (\ref{1a}) that the scalar field equation 
 \be
 \varphi^{|\mu}_{\ |\mu}=-V_{,\varphi}\label{1b}
 \ee
 is satisfied.
 In sections 2-4 we will study black hole solutions of equations (\ref{1a}) and (\ref{1b}) with $\Lambda=0$ and in sections 5-6 we will admit $\Lambda<0$. 
\newtheorem{de}{Definition}
 \begin{de} We say that a $C^2$ solution of (\ref{1a}) and (\ref{1b}) with $\Lambda=0$ is a static spherically symmetric black hole with scalar field if the following conditions are satisfied:
\begin{enumerate}
	\item Metric  $g$ and scalar field $\varphi$ are spherically symmetric and invariant under an additional complete Killing vector $k$.
	\item Metric  is asymptotically flat \cite{w} and $\varphi$ vanishes at infinity.
        \item Vector  $k$ is timelike in the asymptotic region and becomes null on the Killing horizon. Metric and scalar field can be continued through the horizon.
	\end{enumerate}
 \end{de}
 Outside the Killing horizon metric tensor can be written in the form
\be
g=g_{00}dt^2+g_{11}d\rho^2-r^2(d\theta^2+sin^2\theta d\phi^2)\ ,\label{1c}
\ee
where $k=\partial_t$ is the timelike Killing field. Functions $g_{00}$, $g_{11}$, $r$ and $\varphi$ depend only on the coordinate $\rho$.  Coordinates $t,\rho$ break down on the Killing horizon where $g_{00}=0$. Metric can be continued through it if  the Eddington-Finkelstein type coordinates exist. To this end functions $g_{00}$ and $g_{00}g_{11}$ should  have $C^2$ continuations through the horizon and condition
\be
g_{00}g_{11}<0\label{1e}
\ee
should be preserved.
Thus, under the assumptions of Definition 1 functions $g_{00}$, $g_{00}g_{11}$ and $\varphi$ are globally defined and condition (\ref{1e}) has to be  everywhere satisfied.

We use a notion of the asymptotical flatness based on the conformal compactification of spacetime \cite{w}. For our purposes it is sufficient to require that fields have the following  expansions  for large values of $\rho$
\ba \label{1f}
\begin{split}
g_{00}&=1-\frac{2M}{\rho}+0(\rho^{-2}),\ g_{00}g_{11}=-1+0(\rho^{-1}),\\
 r&=\rho+3M+0(\rho^{-1}),\ \ \ \ \ \ \ \ \varphi=0(\rho^{-1})\ .
\end{split}
\ea
The k-derivative of a term $0(\rho^{-n})$, $k,n=1,2$, is assumed to  behave like $0(\rho^{-n-k})$. The constant $M$ is the total mass and the component $3M$ in the relation between $r$ and $\rho$ is introduced for a later convenience. 

Thanks to  (\ref{1e}) one can choose the coordinate $\rho$ in such a way that outside the horizon metric reads
\be
g=Fdt^2-F^{-1}d\rho^2-r^2(d\theta^2+sin^2\theta d\phi^2)\ .\label{1}
\ee
In this gauge one of  the Einstein equations can be easily solved. After doing this we will pass to coordinates $t,r,\theta,\varphi$ which seem more suitable for analyzing solutions.

If $\Lambda=0$ equation (\ref{1a}) for metric (\ref{1}) and scalar field $\varphi$ yields
\be
r^2F_{,\rho\rho}-F(r^2)_{,\rho\rho}+2=0\label{9}
\ee
\be
\varphi'^2=-\frac{2r_{,\rho\rho}}{r}\label{7}
\ee
\be
V=\frac{1}{r^2}-\frac{1}{r^2}(rr_{,\rho}F)_{,\rho}\ .\label{8}
\ee
Above equations are equivalent to equations (9)-(11)  in \cite{bs} for $\bar d=2$ (note that  equation (12) in \cite{bs} contains an error).
By virtue of (\ref{1f}) an integration of (\ref{9}) leads to
\be
r^2F_{,\rho}-F(r^2)_{,\rho}+2\rho=0\label{9c}
\ee
(see equation (13) in \cite{bs}). The second integration allows to represent $F$ in terms of the function $r$
\be 
F=2r^2\int_{\rho}^{\infty}{\frac{\tilde \rho}{r^4}d\tilde \rho}\ .\label{9a}
\ee

It follows from (\ref{7}) that $r_{,\rho\rho}\leq 0$ and from (\ref{1f}) that $r_{,\rho}\ra 1$ if $\rho\ra\infty$.  Hence $r_{,\rho}\geq 1$ everywhere and we can interchange roles of $\rho$ and $r$. In coordinates $t,r,\theta,\varphi$ metric tensor reads
\be
g=Fdt^2-F^{-1}\rho'^2dr^2-r^2(d\theta^2+sin^2\theta d\phi^2)\ ,\label{9b}
\ee
where the prime denotes the derivative with respect to $r$. Now, formulas (\ref{7}), (\ref{8}) and (\ref{9a}) take the following form
\begin{align}\label{27b}
F=2r^2\int_r^{\infty}{\frac{\rho\rho'}{\tilde r^4}d\tilde r}
\end{align}
\begin{align}\label{25b}
\varphi'^2=\frac{2 \rho''}{r\rho'}
\end{align}
\be
V=\frac{1}{r^2}-\frac{1}{r^2\rho'}(\frac{rF}{\rho'})'\ .\label{8a}
\ee
By virtue of (\ref{27b}) the last equation can be also written as 
\ba\label{27d}
V=\frac{1}{r^2}+\frac{2\rho}{r^3\rho'}+\frac{F}{r\rho'^2}(\frac{\rho''}{\rho'}-\frac{3}{r})\ .
\ea

Condition  (\ref{7}) and (\ref{25b}) impose the following simple constraints on function $\rho$
\be
1\geq \rho'>0\ ,\ \ \rho''\geq 0\ .\label{9d}
\ee
In order to guarantee that $F$ and $\varphi$ are  $C^2$ differentiable we have to assume that  $\rho\in C^3$ and 
\be
 (\sqrt{\rho''})'\in C^0\label{19a}
 \ee
at points where $\rho''=0$.  Asymptotic conditions (\ref{1f}) reduce to  
\be 
\rho= r-3M+0(r^{-1})\ \ {\text if}\ \ r\ra\infty\ .\label{26d}
\ee
Note that (\ref{9d}) and   (\ref{26d})   yield 
\be\label{27n}
\rho(r)\geq r-3M\ . 
\ee

In addition to  (\ref{27b})-(\ref{8a}) the scalar field equation (\ref{1b}) has to be taken into account at points where  $\rho''=0$ since then  $\varphi'=0$ at these points.
Let us assume that  $\rho''=0$  at $r=r_0$ which is a limit of  a sequence of points in which $\rho''\neq 0$. Equation (\ref{1b}) will be  satisfied at $r_0$ if values of $V_{,\varphi}=V'/\varphi'$  at these points  have a limit at $r_0$. Here $V'$ is given by differentation of (\ref{27d})
\ba\label{27g}
V'=-\frac{4\rho\rho''}{r^3\rho'^2}+\frac{F}{r^2\rho'^4}(r\rho'\rho'''+7\rho'\rho''-3r\rho''^2)\ .
\ea
Let $S$ be a set of all other points in which  $\rho''=0$. Each of them  possesses a neighborhood where $\varphi=$const. It follows from (\ref{27g}) that also $V=$const in this neighborhood. Equation (\ref{1b}) reduces to  $V_{,\varphi}=0$. It is satisfied provided $V'/\varphi'\rightarrow 0$
when one approaches   boundary of  $S$  from its exterior. Thus, again the continuity of $V_{,\varphi}$ is required. It follows from  (\ref{25b}) and (\ref{27g}) that this property is assured by  the already assumed  condition (\ref{19a}). Concluding, equation (\ref{1b})
does not generate a new constraint.

Given  $V$ as a function of $\varphi$ relations  (\ref{27b})-(\ref{8a}) yield an equation for $\rho(r)$ which is hard to study in an analytic way. For this reason we will   treat $V$ as an unknown variable. Then formulas (\ref{27b})-(\ref{8a})  define $F$, $\varphi$ and $V$ in terms of a free function $\rho(r)$.  We admit only such functions  $\rho(r)$ for which $\varphi(r)$ and $V(r)$ correspond to some potential $V(\varphi)$.

\newtheorem{re}{Remark}
\begin{re}
In principle function $\varphi'$ may change sign at points where $\rho''=0$ or, more generally, when $r$ passes an interval on which $\rho''=0$.   
  If it happens  potential $V$ is a function of $\varphi$ if and only if $\rho(r)$ satisfies in a neighborhood of the interval an involved equation containing  $\rho'$, $\rho''$ and an inverse function to $\rho''/\rho'$.  If there is no interval on which  $\rho$ satisfy this equation  we refer to $\rho$ as a generic function.
\end{re}

If $\rho$ is generic in the above sense it follows from (\ref{25b}) that scalar field is given by the monotonic function
\be 
\varphi(r)=\int^{\infty}_{r}\sqrt{\frac{2 \rho''}{r\rho'}}\label{25d}
\ee
up to the factor $\pm 1$. Formulas (\ref{27d}) and (\ref{25d}) define a differentiable potential $V(\varphi)$ in a parametric way. 

\section{Maximal extension}

Let $r_h$ be a position of the Killing horizon, $F(r_h)=0$. Since $\rho'>0$ everywhere it follows from (\ref{27b}) that $\rho$ cannot have an uniform sign for $r>r_h$. Thus, there is a unique point $r_c$ such that $r_c>r_h$ and
\be
\rho(r_c)=0\ ,\ \ r_c\leq 3M\ . \label{26b}
\ee
Since  $\rho<0$ if $r<r_c$ it is clear from (\ref{27b}) that $(r^{-2}F)'>0$ for $r<r_c$. Hence, there are no more Killing horizons \cite{br1}. 

Using (\ref{27b})  one can calculate the surface gravity of the Killing horizon at $r_h$
\be
\kappa=\frac{|\rho(r_h)|}{r_h^2}\ .
\ee
Thus, the horizon is nondegenerate. Moreover,  its radius   satisfies the Penrose inequality
\be
r_h\leq 2M\ .\label{27h}
\ee
In order to prove (\ref{27h}) let us consider the function 
\be
f=\int_r^{\infty}{\frac{\rho\rho'}{\tilde r^4}d\tilde r} \label{27i}
\ee
which is proportional to $F$ and 
 vanishes at $r_h$.
From (\ref{9d}) and (\ref{27n}) one obtains $\rho\rho'\geq r-3M$  if $r\leq r_c$, hence
\be
f\geq \int_r^{r_c}{\frac{\tilde r-3M}{\tilde r^4}d\tilde r}+f(r_c)\ . \label{27k}
\ee
Integrating (\ref{27i}) by parts yields
\be
f=-\frac{\rho^2}{2r^4}+2\int_r^{\infty}{\frac{\rho^2}{\tilde r^5}d\tilde r}\ . \label{27j}
\ee
Since $\rho(r_c)=0$ it follows   from (\ref{27j}) and (\ref{27n}) that
\be
f(r_c)=2\int_{r_c}^{\infty}{\frac{\rho^2}{\tilde r^5}d\tilde r}\geq 2\int_{3}^{\infty}{\frac{\rho^2}{\tilde r^5}d\tilde r}
\geq 2\int_{3}^{\infty}{\frac{(\tilde r-3M)^2}{\tilde r^5}d\tilde r} =\int_{3}^{\infty}{\frac{\tilde r-3M}{\tilde r^4}d\tilde r}\ . \nonumber
\ee
Substituting the last  relation into (\ref{27k}) and using  (\ref{27b}) leads to the inequality
\be\label{27q}
F(r)\geq 1-\frac{2M}{r}\ \ \texttt{if}\ \  r\leq r_c\ .
\ee 
 Since $r_h<r_c$, condition (\ref{27h}) follows.

If one of  properties (\ref{9d})-(\ref{26d})  breaks down at $r_0$ such that $0<r_0<r_h$ the corresponding field configuration is singular at $r_0$. If not, and $M\neq 0$ a singularity of the Riemann tensor appears at $r=0$ (see the next section). Assume for a moment that $M\leq 0$. It follows from (\ref{27n}) that there is no  zero point $r_c$ of $\rho$. Hence, there is no room for the horizon radius $r_h$. Either $M=0$ and $\rho=r$ (flat space with $\varphi=0$) or there is  naked singularity at $r=0$ or at $r_0>0$.
 
\newtheorem{con}{Conclusion}
\begin{re}
Black hole solutions  can exist  only for $M>0$. 
\end{re}

A question arises what are sufficient conditions for $\rho$ to define  black hole. Of course,
conditions  (\ref{9d})-(\ref{26d}) and  (\ref{26b}) should be satisfied but they do not assure vanishing of  $F$ at some value  $r_h>0$. Assume that these conditions are satisfied   for all  $r>0$.
 In this case, since $\rho(r)$ is increasing and convex,  functions $\rho$  and $\rho'$ are finite at $r=0$ and $\rho(0)<0$. If $\rho'(0)>0$ then it follows from (\ref{27b})  that function $r^{-2}F$ tends to $-\infty$ if $r\ra 0$.  Moreover $F(r_c)>0$ and $(r^{-2}F)'>0$ for $r<r_c$ . Hence, there is exactly one value $r_h$ such that $F(r_h)=0$. The same result follows for a more general condition
on $\rho(r)$ near singularity
\begin{equation}
\rho'\approx cr^n\ ,\ \ c={\text const}>0\ ,\ \ 0\leq n\leq 3\ .\label{29d}
\end{equation}

Let us consider global structure of  the maximally extended black hole solutions. It is convenient to come back to the form  (\ref{1}) of metric tensor.
All  extensions of metrics of this type  were given by Walker \cite{wa} (see also \cite{ch}). They  can be composed from standard building blocks. Since (\ref{27b}) admits  only one horizon and it has the first order zero at $r=r_h$ 
the corresponding maximal extension is obtained by gluing  the same blocks as in the case of the Schwarzschild solution.

We summarize results of this section in  the following theorem. 
\newtheorem{tw}{Theorem}
\begin{tw}
\begin{itemize}
\item A static spherically symmetric black hole with  scalar field corresponds to a function $\rho\in C^3$ with properties (\ref{9d})-(\ref{26d}) and (\ref{26b}). Metric is given by (\ref{9b}) and   (\ref{27b}). Scalar field $\varphi$  and its potential $V$ are given by (\ref{25d}) (or  (\ref{25b}), see Remark 1) and (\ref{27d}).
\item The global structure of the maximally extended metric is the same as that of  the Schwarzschild solution, possibly  with the singularity shifted to  $r_0> 0$. The radius $r_h$ of the horizon  satisfies  
$r_h\leq 2M$ and $r_h<r_c\leq 3M$. The surface gravity is nonzero.
\item Any $C^3$ function $\rho$ satisfying (\ref{9d})-(\ref{26d}), (\ref{26b}) and (\ref{29d}) for all $r>0$ defines a black hole solution  via relations (\ref{9b}), (\ref{27b}), (\ref{27d}) and (\ref{25d}).
\end{itemize}
\end{tw}
\section{Properties of $\varphi$ and $V$ and examples}
First, let us consider  $\varphi$ and $V$ in the region outside the horizon. From (\ref{26d}), (\ref{25b}) and (\ref{27d}) it follows that 
\be
\varphi=0(r^{-1})\ ,\ \ V=0(r^{-3})\label{32g}
\ee
for large values of $r$.
By virtue of (\ref{1b}) and (\ref{32g}) one obtains that potential,  as a function of $\varphi$, satisfies
\be 
V\ra 0\ ,\ \ \ V_{,\varphi}\ra 0\ \ \ {\text if}\ \ \varphi\ra 0\ .\label{32a}
\ee
Somewhere between the horizon and infinity potential $V$ should take negative values. This is the content of  the no-hair theorem of Bekenstein \cite{bek} (with later generalization).

\null

\noindent
{\bf No-hair Theorem.} \emph{Static spherically symmetric black hole with nontrivial scalar field outside the regular horizon cannot have  potential $V$ which is positive definite outside the horizon.}

\null
\noindent
 Following \cite{bs} we show below how this theorem can be proved in our approach. 
One can easily verify that equations (\ref{27b}) and (\ref{27d}) imply the following identity
\ba\label{32d}
2r\rho'V+\rho''(1+\frac{F}{\rho'^2})=(r^2\rho'V -\frac{r\rho''F}{\rho'^2})'    \ .
\ea 
Since  $F(r_h)=0$ and (\ref{32g}) implies $r^2V\ra 0$ if $r\ra \infty$ integration of (\ref{32d}) between $r_h$ and $\infty$ yields
\ba\label{32e}
\int_{r_h}^{\infty}{[ 2r\rho'V+\rho''(1+\frac{F}{\rho'^2})] }= -r_h^2\rho'(r_h)V(r_h)\ .  
\ea 
 If $V\geq 0$ outside the horizon equation (\ref{32e}) is satisfied only if $V=0$ and $\rho''=0$ for every $r\geq r_h$. By virtue of  the asymptotic condition (\ref{26d}) it follows that $\rho=r-3M$ for $r\geq r_h$. Hence, $r_h=2M$ and  one obtains the Schwarzschild metric and trivial scalar field outside the horizon. This solution  may be different  under the horizon since function $\rho=r-3M$ can be continued to $r<2M$ in many different ways respecting  conditions (\ref{9d}) and (\ref{19a}).

The remaining  part of  this section is mainly devoted to behavior  of solutions near  singularity. In order to simplify conclusions we will assume  that $\varphi$ is given by formula (\ref{25d}).
Let singularity be at  $r_0>0$.  Since $\rho$ and $\rho'$ are monotonic and bounded for $r<r_c$ they are finite at $r_0$. Conditions (\ref{9d}), (\ref{19a}) can break down at $r_0$ because $\rho''$ is not well defined at $r_0$ (generic case) or
 $\rho'(r_0)=0$ or $\rho''(r_0)=0$ but $\sqrt{\rho''}$ is not  differentiable at $r_0$. In the first case with $\rho'(r_0)>0$ the scalar field $\varphi$ is finite at $r_0$ and a singular part of  $V$ is  proportional to $-\rho''$. In the case $\rho''(r_0)=\infty$ potential tends to $-\infty$ at $r_0$. Potential of this type is also obtained if  $\rho'(r_0)=0$ and 
$0<\rho''(r_0)<\infty$. In all other cases a dependence of $V$ on $r$ or $\varphi$ is rather not acceptable, e.g. $\varphi$ and $V$ are finite at $r_0$ but the derivative $V_{,\varphi}$ is not defined at $\varphi_0$. We conclude this analysis in the following remark.

\begin{re}
 In generic case  black hole singularity at $r_0>0$ corresponds to a singular potential $V(\varphi)$. If $\rho'>0$ and $\rho''=\infty$  or  $\rho'=0$ and $\rho''>0$ at $r_0$ the scalar field is finite  and $V=-\infty$ at $r_0$.
\end{re}

Let conditions (\ref{9d})-(\ref{19a}) and (\ref{26b})  be satisfied for all $r>0$. Values of $\rho$ and $\rho'$ at $r=0$ will be  denoted by the subscrpt 0. They are finite and they satisfy
\be
-3M\leq \rho_0<0\ ,\ \ 0\leq \rho'_0\leq 1\ .\label{32b}
\ee 
From (\ref{9b}) and (\ref{27b})  one obtains 
\be 
F\approx \frac{2\rho_0\rho'_0}{3r}\ \ \text{if}\ \ r\ra 0\label{33}
\ee
and 
\be
g_{11}(0)=0\ .\label{33b}
\ee
It follows from (\ref{33b}) that the Kretschmann invariant diverges at $r=0$, hence the Riemann tensor is singular. In most cases the function $V(\varphi)$ is also singular. For instance, let us assume that 
$\rho'_0>0$  and $\rho''_0>0$ (generic case). Then $\varphi$ takes a finite value $\varphi_0$ at $r=0$ but $F$ and $V$ tend to infinity. 
Hence, potential $V(\varphi)$ must be singular 
\be 
V\ra \infty\ \ {\text if}\  \ \varphi\ra \varphi_0\ .\label{36}
\ee
An example of this kind is presented in Figure 1 (case 1).

If $\rho'_0>0$ the function $\varphi$ is   finite at $r=0$. Otherwise $\rho''$ would diverge at least as $r^{-1}$, but this  is incompatible with the existence of $\rho'_0$. In order to avoid singularity of $V(\varphi)$ at $\varphi_0$ functions  $V(r)$ and $V'/\varphi'$ should have limits at $r=0$. This implies  condition $r^2V\ra 0$. By virtue of (\ref{27d}) it yields 
\be
(\rho''-3r^2\int_r^{\infty}{\tilde r^{-3}\rho''d\tilde r})\ra 0\ .\label{33a}
\ee
It follows from (\ref{33a}) that $\rho_0''=0$ or $\rho''\sim r^{-1}$. Since the latter condition leads to divergent $\rho'$ we require $\rho_0''=0$.  Still this condition does not guarantee finitness of $V$. For instance, if $\rho'''\ra \rho'''_0\neq 0$  one obtains $V\ra \epsilon \infty$,
 where  $\epsilon=\pm 1$ is the sign of  $\rho'''_0$. To avoid this situation we should   assume  condition $rV\ra 0$ in addition to $\rho_0''=0$. Using (\ref{27d}) yields $\rho'''\sim r^2$. Now $V$ is finite at $r=0$ but to obtain nonsingular $V'/\varphi'$ we still need  stronger conditions near  singularity
 \be
 \rho''\approx c_1 r^5\ ,\ \ (\sqrt{\rho''})'\approx c_2 r^{\frac{3}{2}}\ ,\label{38}
 \ee
 where $c_1$, $c_2$ are nonnegative constants. Note that  condition with $c_2$ can be postponed if $c_1>0$. An example of solution satisfying (\ref{38}) is presented in Figure 1 (case 3).
 
Now, we assume that $\rho_0'=0$. Then $\phi\ra\infty$ if  $r\ra 0$ and potential $V(\varphi)$ is nonsingular.  Generically $V(r)\ra  \pm \infty $  and $V(\varphi)$ behaves like  $\pm \exp{(c\varphi)}$ if $\varphi\ra\infty$, where $c=$const$>0$. An example  of this kind with $\rho_0''>0$ and $\rho_0'''>0$ is presented in Figure 1 (case 2).  In this example function $V(\varphi)$ is positive and proportional to $\exp{(3\varphi/\sqrt{2})}$ for large values of $\varphi$. 

If $\rho_0'=0$ it may happen that function $V(\varphi)$ is bounded. It is the case if $\rho$ admits up to six derivatives at $r=0$  and they satisfy the following conditions 
\begin{align}\label{38a}
\begin{split}
&-3M<\rho_0<0\ ,\ \ \rho''_0>0\ ,\ \ \rho^{(k)}_0=0,\ \ k=1,3,5\\ 
&\rho_0^{(4)}=-\frac{3\rho_0''^2}{\rho_0} ,\ \ \ \ \ \rho_0''^2=4\int_0^\infty{\frac{r\rho_0\rho_0''-\rho\rho'}{r^4}}\ .
\end{split}
\end{align}
We summarize results of this section in the following theorem.
\begin{tw}
If the only singularity is  at $r=0$ then $g_{11}(0)=0$ and the Riemann tensor is singular at $r=0$. Potential  $V(\varphi)$ and its derivative  $V_{,\varphi}$  vanish at $\varphi=0$. Generically $\varphi$ takes a finite value $\varphi_0$ at $r=0$ and $V(\varphi)$ becomes infinite at $\varphi_0$. A nonsingular potential $V(\varphi)$ can be obtained if either  $\rho_0'>0$ and conditions (\ref{38})  are satisfied or $\rho_0'=0$. In the latter case $\varphi$ is unbounded and either $|V(\varphi)|$ grows exponentially or  conditions   (\ref{38a}) are satisfied and $V$ is bounded.
\end{tw}
 Examples of black holes (BH) considered in this section  are given in Figure 1 (cases 1-3). In all of them $M=1$ and $\rho''$ takes the form
\be
\rho''=cr^ne^{-b^2(r-a)^2}\ ,\label{38c}
\ee
where n, a, b and c are parameters. Functions $\rho'$ and $\rho$ are  defined by 
\be
\rho'=1-\int_r^{\infty}\rho''(r')dr'\ ,\label{39a}
\ee
\be
\rho=r-3+\int_r^{\infty}dr'\int_{r'}^{\infty}\rho''(\hat r)d\hat r\ .\label{39b}
\ee
They  can be also expressed without integrals  in terms of functions exp, Erfc and  powers of $r$. Figure 1 contains graphs of functions $\rho$, $g_{00}=F$, $-g_{00}g_{11}=\rho'^2$, $\varphi$, $V$ as functions of $r$ and potential  $V(\varphi)$ as a function of $\varphi$. A zero point of $F$ defines the  radius $r_h$ of the black hole horizon. Potentials $V$ are negative on some intervals in agreement with the no-hair theorems. 

Case 1   represents the generic class of functions $\rho$ for which $\rho'_0>0$, $\rho''_0>0$, $\varphi_0$ is finite and  potential $V$ is infinite at $r=0$. In the case 2 there is $\rho'_0=0$ and $\varphi$ is infinite at $r=0$. Function $V(r)$ is singular at $r=0$ but $V(\varphi)$  is nonsingular and it grows exponentially if $\varphi\ra\infty$. In the case 3 conditions (\ref{38}) are realized. Both $\varphi$ and $V$ are bounded and function $V(\varphi)$  is nonsingular.  Case 4 represents a particle-like solution (to be considered in the next section). Case 5 includes graphs which can be plotted for the Schwarzschild metric.

\begin{figure}[ht]
$\begin{array}{cc}
\includegraphics[width=2.56in]{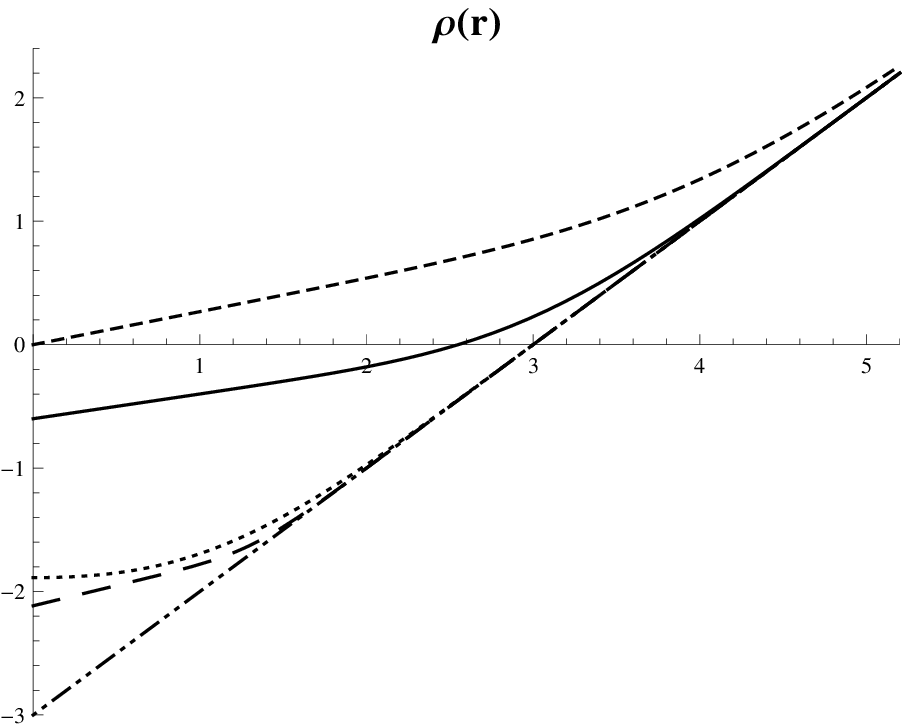} &
\includegraphics[width=2.56in]{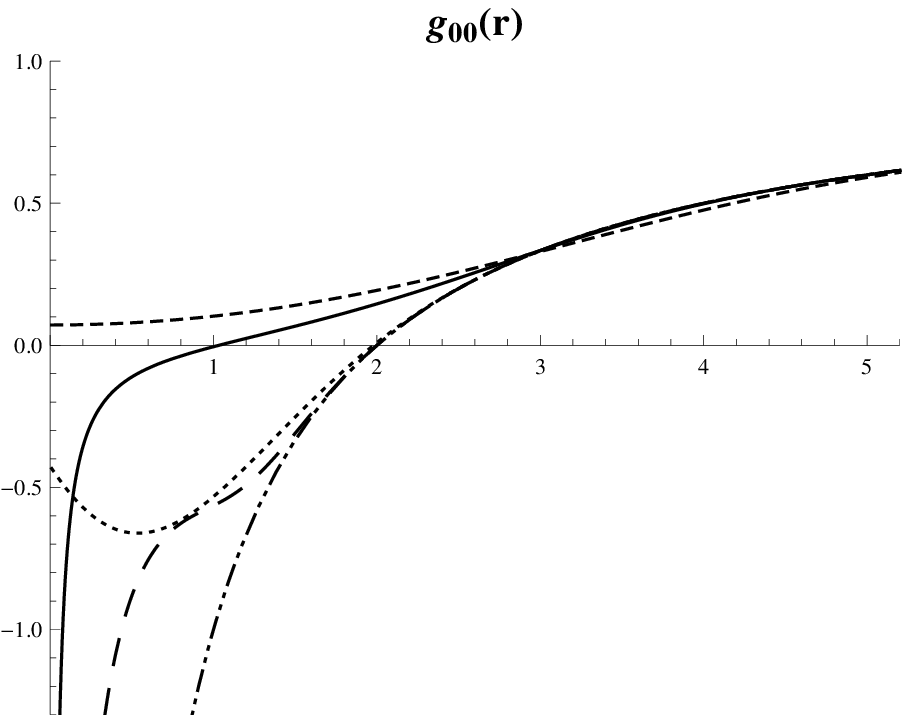}\\
\includegraphics[width=2.56in]{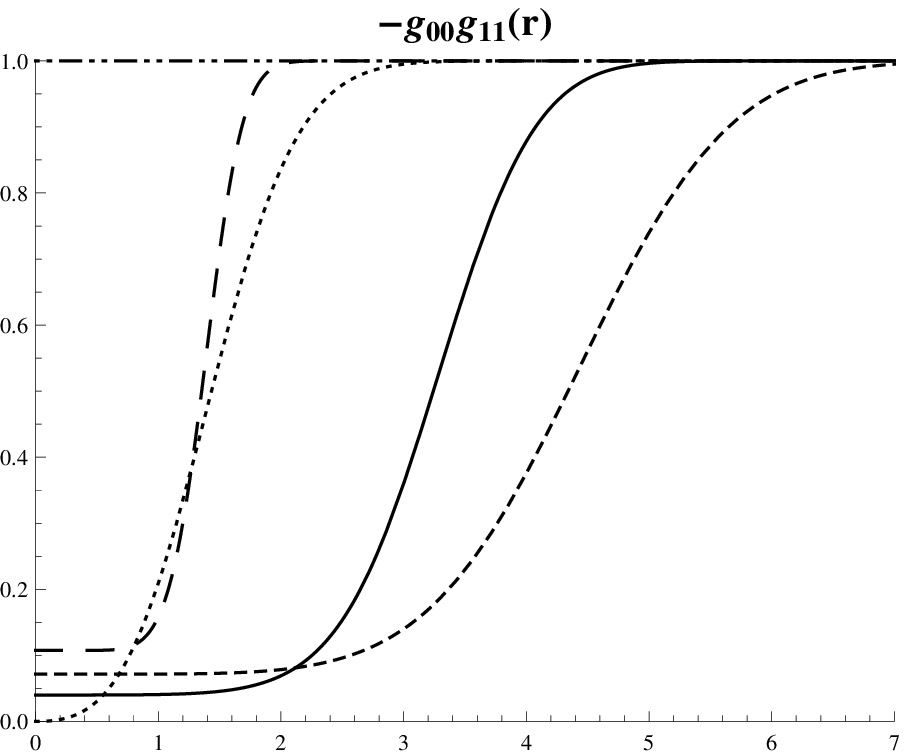} &
\includegraphics[width=2.56in]{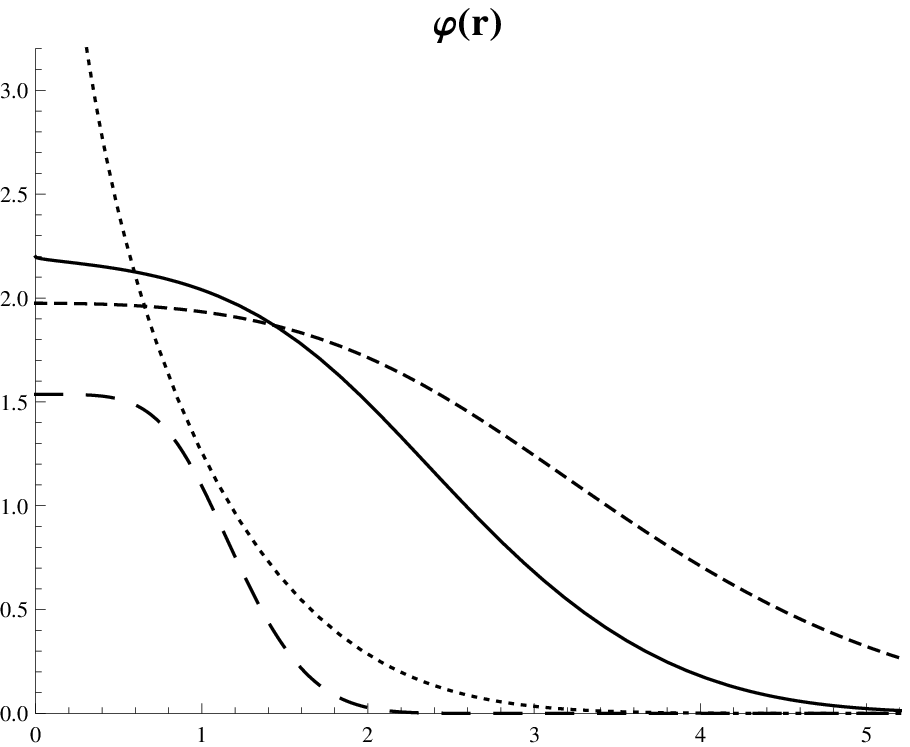} \\
\includegraphics[width=2.56in]{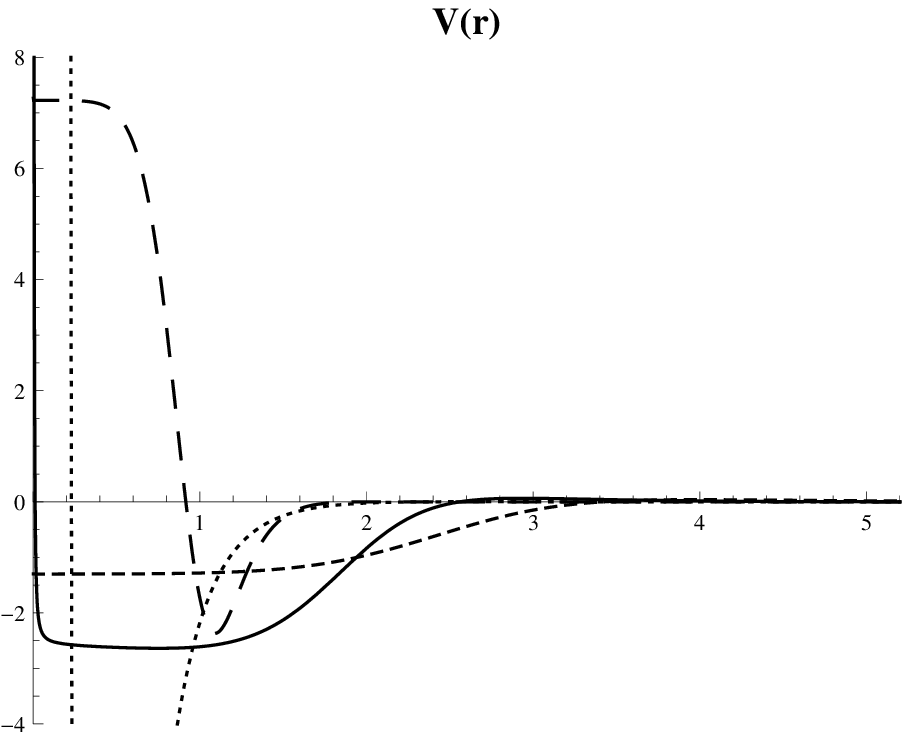} &
\includegraphics[width=2.56in]{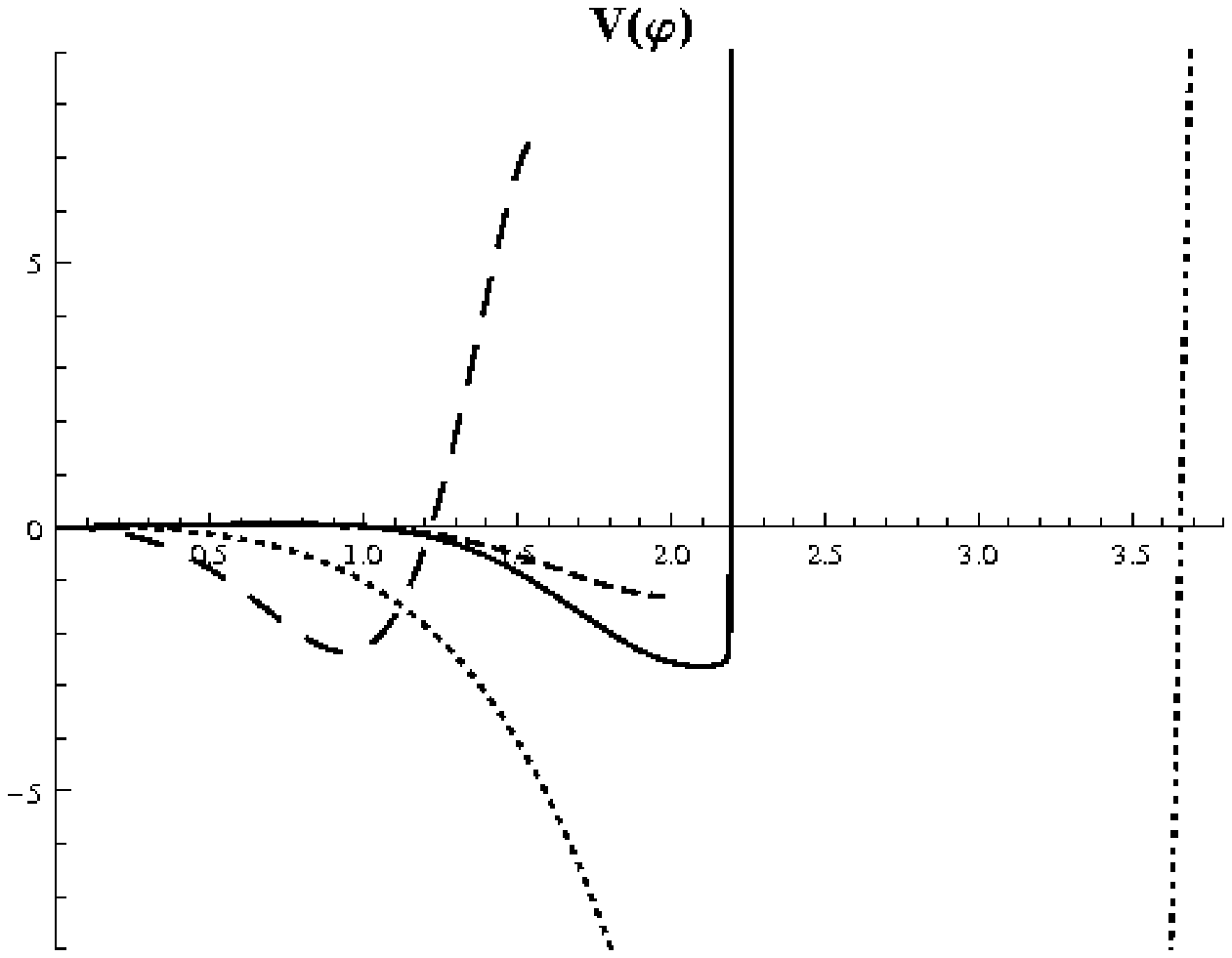} 
\end{array}$
\caption{Solutions corresponding to   (\ref{38c})-(\ref{39b}) with parameters (n,a,b,c) and $\Lambda=0$: 
1 $\frac{\ \ \ \ \ }{\ \ \ \ \ }$ (0,3,1,.45) BH with singular $V(\varphi)$, 
2 $\cdots\cdot\cdot$ (0,1,1,.6) BH with nonsingular $V(\varphi)$ bounded from below, 
3 ---\ --- (5,1,2.5,.5) BH with nonsingular $V(\varphi)$ bounded from both sides, 
4 -\ -\ -\ - (3,3,1,.006) particle like solution with  nonsingular $V(\varphi)$ bounded from both sides, 
5 $\cdot\cdot$--- the Schwarzschild metric}
\end{figure}

 \section{Asymptotically AdS black holes}
 
 Up to now we considered metrics which are asymptotically flat. Not much changes if they are asymptotically anti-de Sitter with the cosmological constant $\Lambda<0$. We will use the term 'AdS black hole with scalar field' if solution is asymptotically AdS and satisfies assumptions of Definition 1 except that about $\Lambda=0$ and asymptotical flatness.  All equations and results of the preceding sections are still true up to the following modifications. Now, instead of (\ref{27b}) one has
\begin{align}\label{46}
F=2r^2\int_r^{\infty}{\frac{\rho\rho'}{\tilde r^4}d\tilde r}-\frac{1}{3}\Lambda r^2\ .
\end{align}
The asymptotic condition (\ref{26d}) is unchanged since  $\rho=r-3M$ is still true for the Schwarzschild-anti-de Sitter solution. Potential $V$ in equations  (\ref{8}), (\ref{8a}) and (\ref{27d}) should be replaced by $V+\Lambda$. Now, equation (\ref{8a}) takes the form
\be
V=\frac{1}{r^2}-\frac{1}{r^2\rho'}(\frac{rF}{\rho'})'-\Lambda\label{51}
\ee
  and equation (\ref{27q}) transforms into
\be\label{45a}
F(r)\geq F_{AdS}=1-\frac{2M}{r}-\frac{1}{3}\Lambda r^2\ \ \ \texttt{if}\ \ r\leq r_c\ .
\ee 
Hence
\be
r_h\leq r_{AdS}\leq 2M\ ,
\ee 
where $r_{AdS}$ is the horizon radius of the AdS metric (zero point of $F_{AdS}$).
  Theorem 1 is replaced by   the following one. 
\begin{tw}
\begin{itemize}
\item
 A static spherically symmetric AdS black hole with  scalar field corresponds to a function $\rho$ with properties (\ref{9d})-(\ref{26d}) and (\ref{26b}). Metric is given by (\ref{9b}) and (\ref{46}). Scalar field $\varphi$  and its potential $V$ are given by  (\ref{25d}) ( or  (\ref{25b}), see Remark 1)  and (\ref{27d}). 
\item The global structure of the maximally extended metric is the same as that of  the Schwarzschild-anti-de Sitter metric, possibly  with a singularity shifted to  $r_0> 0$. A position $r_h$ of the horizon is defined by $F(r_h)=0$ and it satisfies  
$r_h\leq r_{AdS}\leq 2M$ and $r_h<r_c\leq 3M$. The surface gravity is nonzero.
\item Any $C^3$ function $\rho$ satisfying (\ref{9d})-(\ref{26d}), (\ref{26b}) and  (\ref{29d}) for all $r>0$  defines  AdS black hole solution  via relations (\ref{9b}),  (\ref{46}), (\ref{27d}) and (\ref{25d}).
\end{itemize}
\end{tw}
For $\Lambda<0$ estimation (\ref{32g}) changes since potential $V$  vanishes as $0(r^{-2})$ at infinity.  Instead of (\ref{32d}) one obtains
\ba\label{46a}
2r\rho'V+\rho''(1+\frac{F}{\rho'^2}-\Lambda r^2)=(r^2\rho'V -\frac{r\rho''F}{\rho'^2})'    \ .
\ea 
Integrating (\ref{46a}) between $r_h$ and $\infty$ yields
\ba
\begin{split}
&\int_0^{\infty}{[2r\rho'V+\rho''(1+\frac{F}{\rho'^2}-\Lambda r^2)]}=\lim_{r\to\infty}(r^2\rho'V -\frac{r\rho''F}{\rho'^2})-r_h^2\rho'_hV_h=\\
&\Lambda\lim_{r\to\infty}(\frac{r^2}{\rho'}-r^2\rho')-r_h^2\rho'_hV_h\ .\label{50a}
\end{split}
\ea
If $V\geq 0$ outside the horizon then, because of (\ref{9d}), the r. h. s. of (\ref{50a}) is never positive and its l. h. s. cannot be negative. Hence, $V=\rho''=0$ and $\rho'=1$.  It follows that $\rho=r-3M$ if $r\geq r_h$. Thus, $r_h=r_{AdS}$ and for $r\geq r_{AdS}$ scalar field is trivial and metric coincides with the Schwarzschild-AdS solution.  This proves an  analog of Bekenstein's no-hair theorem  for a negative cosmological constant.
\begin{tw}
Static spherically symmetric AdS black hole with nontrivial scalar field outside the regular horizon cannot have  potential $V$ which is positive definite outside the horizon.
\end{tw}
The analysis of solutions near singularity in section 4 applies  to the present case in an unchanged form with the exception of conditions (\ref{38a}) which need  modifications.
\begin{re} 
 Theorem 2 is true for $\Lambda<0$ provided conditions (\ref{38a}) are appropiately changed.
\end{re}
Like for $\Lambda=0$ one  can  construct examples of solutions for $\rho$ given by (\ref{38c})-(\ref{39b}). Corresponding plots are  similar to those in Figure 1 except that now  $F$  is   dominated by the  term proportional to  $\Lambda$  for large values of $r$.

\section{Particle like solutions}

Finally, let us consider particle like solutions which are asymptotically flat or AdS and everywhere regular. In this case the Killing horizon is not admitted since otherwise singularity would be present. Properties (\ref{9d})-(\ref{26d}) and $F>0$ have to be satisfied for all $r>0$. Inspection of the Kretschmann invariant at $r=0$ shows  that   metric should  flatten at $r=0$. It follows that
 \be
 0<F_0<\infty\ ,\ \ \frac{\rho_0'^2}{F_0}=-1\ .\label{39}
 \ee
This together with (\ref{33})  yields 
\be 
\rho_0=0\ ,\ \ \rho_0'>0\ .\label{40}
\ee
In order to avoid singularities of the first and second derivatives of metric  with respect to the Cartesian coordinates related in the standard  way to  $r$, $\theta$ and $\phi$ we should also assume that 
\be
\rho_0''=\rho_0'''=0\ .\label{41}
\ee  
To obtain  $\varphi\in C^2$ and a nonsingular function $V(\varphi)$ we still  have to strenghten (\ref{41}) by assuming
\be
\rho''\approx 4cr^3\ ,\ (\sqrt{\rho''})'\approx 3c\sqrt{r}\ ,\ c=\texttt{const}>0\ \ \text{if}\ \ r\ra 0\ .\label{44}
\ee
\begin{tw}
 \begin{itemize}
\item A static spherically symmetric, asymptotically flat or AdS,  everywhere regular solution with  scalar field corresponds to some function $\rho$ with properties (\ref{9d})-(\ref{26d}), (\ref{40}) and (\ref{44}). Metric is given by (\ref{9b}) and either (\ref{27b}) or (\ref{46}). There is  no horizon.  Scalar field  and its potential $V$ satisfy (\ref{25d}) (or  (\ref{25b}), see Conclusion 2)  and (\ref{27d}). 
\item Any $C^3$ function $\rho$ satisfying (\ref{9d})-(\ref{26d}), (\ref{40}) and (\ref{44}) for all $r> 0$ defines a particle like solution  via relations   (\ref{9b}),  (\ref{27d}), (\ref{25d}) and either (\ref{27b}) or (\ref{46}).
\end{itemize}
\end{tw}

If $\rho\in C^5$ 
 conditions (\ref{40}) and (\ref{44}) are satisfied  by
 \be
 \rho=c_1r+c_2r^5f\ ,\label{44a}
 \ee
 where $c_1$ and $c_2$ are positive constants and $f$ is a function such that $f\ra 1$ if $r\ra 0$. An asymptotically flat example  of this kind is represented by the  case 4 in Figure 1. The corresponding function $V(\varphi)$  is defined only on an interval and it can be prolongated in any differentiable way.

Following \cite{bs}, where the flat asymptotic was considered, we can show for any $\Lambda\leq 0$ that potential $V$ has to be somewhere negative if $\varphi$ is nontrivial. Indeed, integrating of (\ref{51}) from zero to infinity yields
\ba
\begin{split}
&\int_0^{\infty}{r^2\rho'Vdr}=\lim_{r\to\infty}(\rho-\frac{rF}{\rho'}-\Lambda\int_r^{\infty}{\tilde r^2\rho'd\tilde r})=\\
&-M+\frac{1}{3}\Lambda\lim_{r\to\infty}(\frac{r^3}{\rho'}-r^3\rho'+\int_r^{\infty}{\tilde r^3\rho''d\tilde r})\ .\label{50}
\end{split}
\ea
Because of  (\ref{9d}) the r. h. s. of (\ref{50}) is never positive. If $V\geq 0$ then it follows from this equation that $V=M=0$ and $\rho=r$. Hence, scalar field is trivial and metric is flat or AdS. In this way we obtain the following generalization of Theorem 5 in \cite{bs}.
\begin{tw}Static spherically symmetric, asymptotically flat or AdS,  everywhere regular solution with nontrivial  scalar field
cannot have positive definite potential $V$.
\end{tw}
If $M<0$ it follows from (\ref{27n})  that condition  (\ref{40}) cannot be satisfied. For $M=0$ the only nonsingular configuration (flat or AdS metric with $\varphi=V=0$) is given by $\rho=r$. As in the case of black holes  nontrivial particle like solutions exist only if $M>0$. This observation together with Remark 1 leads to the following statement.
\begin{tw}
For $M\leq 0$ there are no  static spherically symmetric, asymptotically flat or AdS, black holes or particle like solutions  with nontrivial scalar field.
\end{tw}
Thus, in case of these solutions one can assume $M>0$ without loss of generality.
\section{Summary}
We have been studying static spherically symmetric  solutions of the Einstein-scalar equations which are asymptotically flat or AdS and have no naked singularities. Our approach is based on a representation of these solutions in terms of a function $\rho(r)$ which is monotonically growing, convex and tends to $r-3M$ if $r\ra\infty$.  Moreover it should have a zero point in the case of black holes or to satisfy conditions (\ref{40}) and (\ref{44}) in the case of particle like solutions. Necessary and sufficient conditions for the function $\rho$ are given in Theorems 1, 3 and 5.

Theorem 7 shows that the total mass of field configurations with nontrivial scalar field has to be positive. This particular version of the positive energy theorem  is satisfied for any shape of potential $V(\varphi)$. Note that general $V(\varphi)$  does not assure any reasonable energy condition except the null one. 

Theorems 4 and 6 generalize to the AdS asymptotic the no-go theorems for  asymptotically flat black holes and particle like solutions. They show that  potential $V$ has to be somewhere negative or $V=\varphi=0$ everywhere. In agreement with \cite{br1} the global structure of solutions is  that of the Schwarzschild or Minkowski metric or their AdS partners (Theorems 1 and 3). We also show that in the case of black holes horizons are not degenerate (the surface gravity is nonzero) and the Penrose inequality $r_h\leq 2M$ or stronger $r_h\leq r_{AdS}$ is satisfied (Theorem 1 and 3).

In our approach an exact form of the function $V(\varphi)$ is not a'priori assumed. It follows implicitly when  function $\rho(r)$ is 
chosen. Some  properties of $V(\varphi)$, like regularity or boundedness, can be easily related to behavior of $\rho$ near singularity. For generic $\rho$ potential $V(\varphi)$ becomes infinite at a finite value of $\varphi$. The simplest way to make it regular  in the case of black holes is to assume that the first  derivative of $\rho$ at $r=0$ vanishes. Then $V$ is bounded from below or above and $|V|$ grows exponentially when $\varphi\rightarrow\infty$. To get potential bounded from both sides one has to assume stronger conditions (\ref{38}) or (\ref{38a}) (modified  in the case of the AdS asymptotic). In the first case the function $\varphi(r)$ is also bounded. Hence, potential $V(\varphi)$ is given only on an interval and it can be prolongated in any way. Similar situation takes place in the case of  particle like solutions. Properties of $V(\varphi)$ for black holes are summarized in Theorem 2 and Remark 3.

Thanks to sufficient conditions on $\rho$ given in Theorems 1, 3 and 5  it is easy to construct examples of different classes of solutions. Some of them, with a singular or nonsingular potential $V(\varphi)$, are presented in Figure 1. Plots were done by means of  Mathematica 9.

\section*{Acknowledgments}
I am grateful to Piotr Chru\'sciel for discussions and an access to his lecture notes \cite{ch}. 
This work is partially supported by the grant N N202 104838 of Ministry of Science and Higher Education of Poland.


\begin{thebibliography}{99} 
\bibitem{f}
Fisher I.Z.: Scalar mesostatic field with regard for gravitational effects, Zh. Eksp. Teor. Fiz.
\textbf{18}, 636-640 (1948), English translation: gr-qc/9911008
\bibitem{s}
 Szekeres G.: New Formulation of the General Theory of Relativity, Phys. Rev. {\bf 97},  212 (1955)
\bibitem{bl}
 Bergmann O.,  Leipnik R.: Space-time structure of a static spherically symmetric scalar field,
Phys. Rev. \textbf{107}, 1157 (1957)
\bibitem{b}
Buchdahl H.A.: Reciprocal static metrics and scalar fields in the general theory of relativity, Phys. Rev. {\bf 115}, 1325 (1959)
\bibitem{jnw}
Janis A.I.,  Newman E.T.,  Winicour J.: Relativity of the Schwarzchild singularity, Phys. Rev. Lett. \textbf{20}, 878 (1968)
\bibitem{r}
Rosen G.: Existence of Particlelike Solutions to Nonlinear Field Theories, J. Math. Phys. \textbf{ 7},  2066 (1966)
\bibitem{cha}
Chase J.E.: Event Horizons in Static Scalar-Vacuum Space-Times, Commun. Math. Phys. \textbf{19},  276 (1970)
\bibitem{bek}
Bekenstein J.D.: Nonexistence of Baryon Number for Static Black Holes, Phys. Rev. \textbf{D 5},  1239 (1972)
\bibitem{tei}
Teitelboim C.: Nonmeasurability of the Baryon Number of a Black-Hole, Lett. Nuovo Cimento \textbf{3},  326 (1972)
\bibitem{ap}
Adler S.L.,  Pearson R.B.: "No-hair" theorems for the Abelian Higgs and Goldstone model, Phys. Rev. \textbf{D 18}, 2798 (1978)
\bibitem{h}
Heusler M.: A no-hair theorem for self-gravitating nonlinear sigma models, J. Math. Phys. \textbf{33},  3497 (1992)
\bibitem{sud}
Sudarsky D.: A simple proof of a no-hair theorem in Einstein-Higgs theory, Class. Quantum Grav. \textbf{12},  579 (1995)
\bibitem{bek2}
Bekenstein J.D., Novel ``no-scalar-hair`` theorem for black holes, Phys. Rev. \textbf{D 51},  R6608 (1995)
\bibitem{gl}
Gal'tsov D.V., Lemos J.P.S.: No-go theorem for false vacuum black holes, Class. Quantum Grav. \textbf{18 },  1715 (2001)
\bibitem{br1}
Bronnikov K.A.: Spherically symmetric false vacuum: No-go theorems and global structure
Phys. Rev. \textbf{D 64}, 064013 (2001)
\bibitem{bs}
Bronnikov K.A., Shikin G.N.: Spherically symmetric scalar vacuum: no-go theorems, black
holes and solitons, Grav. Cosmol. \textbf{8},  107 (2002)
\bibitem{br}
Bronnikov K.A., Rubin S.G.: \textit{Black Holes, Cosmology and Extra Dimensions}, World Scientific (2012)
\bibitem{tmn}
Torii T.,  Maeda K., Narita M.: Scalar hair on the black hole in asymptotically anti-de Sitter spacetime, Phys. Rev. \textbf{D 64},  044007 (2001)
\bibitem{ns}
Nucamendi U., Salgado M.: Scalar hairy black holes and solitons in asymptotically flat spacetimes,  Phys. Rev. \textbf{D 68 },  044026 (2003)
\bibitem{mtz}
Martınez C., Troncoso R., Zanelli J.: Exact black hole solution with a minimally coupled scalar field, Phys. Rev. \textbf{D 70 },  084035 (2004)
\bibitem{ao}
Anabalon A., Oliva J.: Exact Hairy Black Holes and their Modification to the Universal Law of Gravitation, Phys.Rev. \textbf{D86}, 107501 (2012)
\bibitem{w}
Wald R.: General Relativity, University of Chicago Press (1984)
\bibitem{wa}
Walker M.: Bloc diagrams and the extension of timelike two–surfaces, J. Math. Phys. \textbf{11}, 2280–2286 (1970)
\bibitem{ch}
Chru\'sciel P.: Lectures on Black Holes, Krak´ow  (2009)
\end{thebibliography}
\end{document}